\documentclass{article}
\usepackage{amsmath,amsfonts,amsthm,amssymb,amscd,color,xcolor,mathrsfs}
\usepackage{amsfonts}
\usepackage{amsmath,amsthm}
\usepackage{hyperref}
\usepackage{latexsym}
\usepackage{array}
\usepackage{amssymb}
\usepackage{enumerate}

\usepackage[francais,english]{babel}

\usepackage{color}
\usepackage[latin1]{inputenc}

\DeclareFontFamily{U}{wncy}{}
    \DeclareFontShape{U}{wncy}{m}{n}{<->wncyr10}{}
    \DeclareSymbolFont{mcy}{U}{wncy}{m}{n}
    \DeclareMathSymbol{\Sh}{\mathord}{mcy}{"58}

\theoremstyle{plain}
\newtheorem{theorem}{Theorem}[section]
\newtheorem*{theorem*}{Theorem}

\theoremstyle{remark}

\newtheorem*{lem*}{Lemma}
\newtheorem*{sublem*}{Sublemma}
\newtheorem*{remark*}{Remark}
\newtheorem*{NB*}{NB}

\newcommand{\PP}{ \mathbb{P}\,}

\newcommand{\R}{ \mathbb{R} }

\newcommand{\Z}{ \mathbb{Z} }
\newcommand{\N}{ \mathbb{N} }

\newcommand{\T}{ \mathbb{T} }

\newcommand{\cD}{ \mathcal{D} }

\newcommand{\cF}{ \mathcal{F} }

\newcommand{\EE}{ {\mathbb E}}

\newcommand{\om}{ \omega }

\renewcommand{\phi}{ \varphi }

\newcommand{\gsim}{ \gtrsim }

\newcommand{\be}{\begin{equation}}
\newcommand{\ee}{\end{equation}}
\newcommand{\ben}{\begin{equation*}}
\newcommand{\een}{\end{equation*}}

\newcommand{\bt}{\begin{theorem}}
\newcommand{\et}{\end{theorem}}

\numberwithin{equation}{section}

\newcommand{\llan}{ \langle\!\langle }
\newcommand{\lann}{ \langle\!\langle }

\newcommand{\rran}{ \rangle\!\rangle}
\newcommand{\rann}{ \rangle\!\rangle}

\newcommand{\p}{ \partial}

\author{Sergei Kuksin\footnote
{{S. Kuksin, Institut de Math\'emathiques de Jussieu-Paris Rive Gauche, CNRS, Universit\'e Paris Diderot, UMR 7586, Sorbonne Paris Cit\'e, F-75013, Paris, France; and
School of Mathematics, Shandong University, Jinan, 250100, PRC; and
Saint Petersburg State University, Universitetskaya nab., St. Petersburg, Russia,}
 e-mail: \href{mailto:Sergei.Kuksin@imj-prg.fr}{sergei.kuksin@imj-prg.fr}}
}

\title
{  Kolmogorov's theory of turbulence and  its rigorous 1d model. }

\begin{document}
\date{}

%\centerline {\today}
\maketitle

\hfill{\it Dedicated to my friend Alexander Shnirelman  } 

\hfill{\it  on his  75-th
birthday } \medskip

\tableofcontents

\begin{abstract}
This paper is a synopsis  of the recent book \cite{BK}. 
% A. Boritchev, S. Kuksin, \textit{One-Dimensional Turbulence and  the Stochastic Burgers Equation}, MS of a book, 2020. 
The latter  is dedicated to the stochastic Burgers equation as a model for 1d turbulence, and the paper discusses its content 
in  relation to the Kolmogorov theory of turbulence. 
 \end{abstract}

 \centerline{\bf Résumé}
 Cet article est un synopsis du livre récent \cite {BK}. Le livre est dédié à l'équation be Burgers stochastique comme un modèle du turbulence 
 unidimensionnelle, et l'article discute de son contenu en relation avec la théorie de la turbulence de Kolmogorov.
 % et l'article  discute de son contenu par rapport à la  théorie de la turbulence de Kolmogorov.
%\medskip 

\section{Introduction}

The goal of this  paper is to discuss the content  of the book \cite{BK}, dedicated to a rigorous theory of 1d turbulence, in its
 relation to Kolmogorov's  understanding   of hydrodynamical   turbulence, known as the K41 theory. 
 At the origin of the book lie the results, obtained in PhD theses of two students of the author of 
 this paper,  A.~Biryuk  \cite{BirPhD,Bir01} and A.~Boritchev  \cite{BorPhD, BorW} (the latter is another  author of the book \cite{BK}). The  theses, in their turn, 
 were   based on the previous work 
 \cite{K97, K99Tr, K99G}  on turbulence in the complex Ginzburg--Landau  equation (see \cite[Section~5]{BJPV} for this concept). The results of the two theses 
 were developed further  in  subsequent publications of their authors, 
  gave a material for an M2 lecture course which the author of this paper taught in Paris~7 and in some other 
 universities, and were improved and edited in the lecture notes for that  course \cite{KK}.  Finally  the results were significantly  developed 
 while working on the book \cite{BK}. More detailed references may be found in Chapters~5 and 9 of  \cite{BK}. Proofs of all theorems, given below,
 may be easily found in \cite{BK}. Some  of them are sketched in the paper.

The paper is based on a number of zoom-seminars which we gave in the year 2020. 
 
\subsection{K41 theory}

The K41 theory of turbulence was created by A.~N.~Kolmogorov in three articles \cite{Kol41a, Kol41b, Kol41c},  published in 
1941 (partially based on the previous work of  Taylor and von~Karman--Howard); see in 
 \cite[$\S{33\!-\!34}$]{LL}, \cite{ F} and 
\cite[Chapter~6]{BK}.  This heuristic theory 
 describes statistical properties of turbulent flows of fluids and gazes 
 and is now the most popular theory of turbulence. We will discuss its basic concepts for the case of 
 a fluid flow with  velocity  $u(t,x)$  of order 1, space--periodic of period one and with zero space-meanvalue. 
 The  Reynolds number  of such a  flow   is 
\be\label{Rey}
Rey % = \frac{\langle\text{space-scale of the flow} \rangle \langle\text{characteristic speed of the flow} \rangle }
= \nu^{-1},
\ee
where $\nu$ is the fluid's 
 viscosity. If $Rey=\nu^{-1}$ is large, then the velocity field  $u(t,x)$  
becomes very irregular, and the flow becomes 
 turbulent. The viscosity  is the most important parameter for what follows; dependence on it is clearly indicated, 
and all constants below are  independent from $\nu$. 

Kolmogorov postulated  that the  short  scale in $x$ features   of a turbulent flow $u(t,x)$ 
 display a universal behaviour which  depends on  particularities of the system only through a few parameters (in  our setting 
 -- only through $\nu$), and the K41 theory presents and discusses these universal features -- the laws of the Kolmogorov theory.

  The K41 theory is statistical. That is, it assumes  that the velocity    $u(t,x)$ is a random field over some probability space 
    $
   (\Omega, \cF, \PP).
  $
  Moreover, %Kolmogorov   postulates  that 
   $u$ is  assumed to be stationary in time
   and homogeneous in space with zero mean-value,
   $ \ 
   \EE u(t,x) \equiv 0.
   $
     The K41 theory studies its short space-increments 
  $ \ 
  u(t, x+r) - u(t,x)$, $ |r|\ll 1, 
  $
  and examines their moments as functions of $r$. Besides, for the decomposition of  $u(t,x)$ in Fourier series 
  $$
  u(t,x) = \sum_{s \in \Z^3 } \hat u_s(t) e^{2\pi i s\cdot x}, \quad \hat u_0(t) \equiv0, 
  $$
  the theory  examines  the second moments of   Fourier coefficients $ \hat u_s(t)$ as functions of $|s|$ and $\nu$.

  Below we present the one-dimensional version of the Kolmogorov theory for a model, given by the stochastic Burgers equation and 
  advocated by Burgers, Frisch, Sinai and some other mathematicians and physicists. Then we will discuss the basic statements of the
  K41 theory, their 1d versions and the proofs of the latter, suggested in \cite{BK}.

\subsection{Stochastic Burgers equation} 
 
 The model for 1d turbulence we will talk about % suggests as a model for 1d turbulence (i.e. for burgulence)
 is given by   the stochastic  Burgers equation 
  \begin{equation} \label{B}
 \begin{split}
u_{t} +uu_{x} -\nu u_{xx} = \p_t \xi(t,x),\quad\   x\in S^1=\R/\Z, 
 \;\; \int u\,dx=\!\!\int \xi\,dx=0, \\
u(0,x)=u_0(x),
% x\in S^1,\;\;\; \int u\,dx=0,\;\; 0<\nu\le1, 
 \end{split}
\end{equation}
where  $\xi$ is a Wiener process in the space of functions of $x$,
 \be\label{force}
\xi^\omega(t,x) = \sum_{s= \pm1, \pm2, \dots} b_s \beta_s^\omega (t) e_s(x) ,\qquad 
0<B_0 = \sum_s b_s^2<\infty. 
\ee
Here $\{ e_s, s=\pm1, \pm2, \dots\}$ is the trigonometric basis in the space of 1-periodic function with zero mean:
\[
\left\{
\begin{array}
[c]{c}
e_{k}=\sqrt{2}\cos(2\pi kx),\\
e_{-k}=\sqrt{2}\sin(2\pi kx),
\end{array}
\right. \;\; k \in
\N  ;
\]
 $\{ \beta_s^\omega (t)\}$ are 
standard independent Brownian processes, and $\{ b_s \}$ are real numbers. For the purposes on this paper we assume that they 
fast converge to zero. Then 
$\xi$ is a Wiener process in the space of functions of $x$, and for a.e. $\om$ 
its realisation $\xi^\omega(t,x)$ is continuous in $t$ and  smooth in $x$.

As usual, $u^\om(t,x)$ is a solution of \eqref{B} if a.s. 
 $$
u(t)- u_0 +\int_0^t (uu_x -\nu u_{xx}) \, ds = \xi(t), \qquad \forall\, t\ge0.
$$
For an integer $m\ge0$ we denote by $H^m$ the $L_2$--Sobolev space of order $m$ of functions on $S^1$ with zero mean-value, equipped 
with the homogeneous norm
$$
\| u\|_m^2 = \int (  u^{(m)}(x))^2 dx.
$$
 It is not  hard to see that if $u_0\in H^r,\ r\ge1$, 
 then  there is a  solution $u$ of \eqref{B} such that   $ u^\om  \in C(\R_+, H^r)$  a.s., and any two solutions coincide  a.s. 
  We will denote a solution of \eqref{B}, 
 regarded as a random process in a space of functions of $x$  as  $u(t;u_0)$ or $u^\nu(t;u_0)$. Regarding $u$ as a random field of $(t,x)$ we
 will write it as $u(t,x;u_0)$ or $u^\nu(t,x;u_0)$.

As we will soon explain, in average   solutions   of \eqref{B} are  of order one, i.e. for 
  any $u_0$,  \ $\EE | u^\nu (t;u_0)|^2_{L_2} \sim1  $ uniformly in $t\ge 1$ and 
  $\nu\in(0,1]$. 
   Since the order of magnitude of a solution   $u^\nu$ is  $\sqrt{ \EE | u^\nu(t)|^2_{L_2} }$ and its 
    space-period is one, then  the     Reynolds number of  $u^\nu$   is 
   $
  % \text{Rey}(u^\nu) 
    \sim \nu^{-1},
   $
   as in \eqref{Rey}. 
So  eq.~\eqref{B} with small $\nu$ describes  1d turbulence (called by Uriel~Frisch {\it burgulence}). 
\medskip

The goals, related to eq. \eqref{B} as a 1d model of turbulence, are:\\
1) to study  solutions $u^{\nu }(t,x)$  for small $\nu$ and for $1\le  t \le\infty$;\\
2) 
to relate the obtained results with the theory of turbulence, 
regarding the Burgers equation \eqref{B} as a 1d hydrodynamical equation.

Inspired by the heuristic work on the stochastic Burgers equation by U.~Frisch with collaborators (e.g. see \cite{Fr1, Fr2}), 
 Sinai  and others in the influential   paper \cite{Sin} used the Lax-Oleinik formula to write down 
the limiting dynamics of \eqref{B} as $\nu\to 0$,  and next studied  the obtained 
limiting  solutions  $u^0(t,x)$ of the inviscid stochastic Burgers equation \eqref{B}${}\!\mid_{\nu=0}$.  The research was continued by Khanin and 
some other mathematicians, e.g. see  \cite{Kha} and 
references in \cite{BK}. It has led to  a beautiful theory which is related to 1d turbulence and casts light on  the problem~1) above, but so far this 
 approach has  not allowed   to obtain for the limiting dynamics  analogies of the K41 laws.

On the contrary,  in \cite{BK} we study eq.~\eqref{B} for small but positive $\nu$, i.e. not when
  $\nu\to0$, but when  $0<\nu\ll1$ is fixed, using basic tools from PDEs and stochastic calculus. This approach 
 allows to get  relations, similar to those claimed by the K41 theory, and to
 rigorously justify the heuristic theory of burgulence,  built in \cite{Fr1, Fr2}.

\medskip

\noindent 
{\bf Acknowledgments.}  The author was  supported by  the grant  18-11-00032 of Russian Science Foundation.

\section{Apriori estimates}\label{s_2}
 We start with  a-priori estimates for equation  \eqref{B}. 
 The key starting point is the Oleinik  inequality, which we apply to solutions of \eqref{B} with  fixed $\omega$. 
The inequality was proved by Oleinik for the free Burgers equation, but   her argument  applies to the stochastic equation \eqref{B} trajectory-wise  and  implies the following result:

\bt\label{t_olei}
For any initial data $u_0\in H^1$, any $p\ge1$
  and any $\nu, \theta\in (0,1]$,  uniformly in $t\ge\theta$ 
   we have:
   \begin{equation} \label{OK}
  \EE %\sup_{\theta \le t \le \theta+ T} 
  \big(
  %|u_x^{\nu+}(t,\cdot)|_\infty^p  + 
   |u^\nu(t;u_0)|_{L_\infty}^p+  |u^\nu_x(t;u_0)|_{L_1}^p  +| u_x^+(t;u_0)|_{L_\infty}^p
   \big)  \le C_p  \theta^{-p}
  \ee
    (here $v^+=\max(v,0)$).  Apart from $p$, the   constant $C_p$ depends only on the random force in \eqref{B}.
    \et
    
    Decomposing a solution $u^\nu(t;u_0)$ in Fourier series,
  \be\label{F}
    u^\nu(t,x;u_0) =\sum \hat u_k^\nu(t;u_0) e^{2\pi ikx},
   \ee
   and using that $| \hat u_k^\nu(t;u_0) | \le | u^\nu_x(t;u_0)|_{L_1} / 2\pi |k|$ we derive from \eqref{OK} an important consequence:
   \be\label{conseq}
   \EE | \hat u_k^\nu(t;u_0) |^p \le C'_p |k|^{-p} \theta^{-p}, \quad p\ge1,\; \; |k| \ge1, 
   \ee
   if $t\ge\theta$, for any $u_0\in H^1$.

\subsection{Upper bounds for moments of Sobolev norms of solutions} 

The very powerful estimate \eqref{OK},  jointly with some PDE tricks, allows to bound  from above  moments of all Sobolev norms of solutions. Namely,
denoting 
 $$
 X_{j}(t)=\mathbb{E} \|{u(t)}\|_{j}^{2},  \quad B_{m}=\underset{s\in
\mathbb{Z}^{\ast}}{\sum}|2\pi s|^{2m}b_{s}^{2} < \infty\,,
 \qquad j,m\in\N,
 $$
 and applying to eq. \eqref{B} Ito's formula, estimate 
 \eqref{OK}  and the Gagliardo-Nirenberg inequality we get that 
\[
\begin{split}
\frac{d}{dt}X_{m}(t)\leq &\ B_{m}-2\nu X_{m+1}(t)+C_{m}
X_{m+1}(t)^{\frac{2m}{2m+1}}\\
= &\ B_m-X_{m+1}(t)^{\frac{2m}{2m+1}}\left(
2\nu X_{m+1}(t)^{\frac{1}{2m+1}}-C_{m}\right), 
\quad t\ge\theta.
\end{split}
\]
Using once again \eqref{OK} jointly with basic PDE inequalities we obtain
\[
X_{m}(t)  %C_{m1}X_{m+1}(t)^{\frac{2m-1}{2m+1}}\left(  \mathbb{E[}|u(t)|_{1,1}^{a}]\right) ^{b}
\leq C'_{m}X_{m+1}(t)^{\frac{2m-1}{2m+1}},  \quad t\ge \theta.
\]
It can be derived from these  two relations  that second moments of
 $L_2$-Sobolev norms of solutions are bounded uniformly in the initial data:

\bt\label{main_est}
  For any $u_0\in H^1$, every $m\in\N$, $0<\nu\le1$  and every $\theta>0$, 
  \be\label{upper}
  \EE \| u^\nu(t;u_0)\|_m^2 \le C(m, \theta) \nu^{-(2m-1)} \quad \text{if} \quad  t\ge \theta. 
  \ee
  \et
  
  Jointly with \eqref{OK} and the Gagliardo-Nirenberg inequality this result implies upper bounds on moments of all $L_p$--Sobolev norms of 
  solutions of \eqref{B}. A remarkable feature of the Burgers equation is that these  estimates  % for  moments of the  $L_1$--Sobolev norms.}
  are asymptotically sharp when $\nu\to0$.\footnote{Except the estimates 
     with $p=1$, when we do not know if this is true.}
   In the next section we prove this fact for the basic inequalities   \eqref{upper}.

\subsection{Lower bounds} 
 
 The Ito formula, applied to $\frac12 \| u(t)\|_0^2$, where $u(t)$ satisfies \eqref{B}, implies the  balance of energy relation 
 $$
 {\mathbb E} \int \tfrac12 |u(T+\sigma, x)|^2dx -  {\mathbb E} \int\tfrac12 |u(T, x)|^2dx  + \nu {\mathbb E} \int_T^{T+\sigma} \!\!\!\int |u_x(s,x)|^2dxds =  \sigma B_0,
$$
where $T, \sigma >0$.
Let $T\ge1$. By  \eqref{OK} the first two terms are bounded by a constant $C_*$ which depends 
only on the random force. If $ \sigma \ge \sigma_* = 4C_*/B_0$, then  $C_*\le \tfrac14 \sigma B_0$ and   we get that 
  
 $$
 \nu {\mathbb E}\, \frac1{\sigma}  \int_T^{T+\sigma} \!\!\!\int |u_x(s,x)|^2dxds \ge \tfrac12 B_0.
 %4\,\frac{C_*}{\sigma}.  
$$

For any 
random process $f^\om(t)$ we denote by $\lann f \rann $ its averaging  in ensemble and local  averaging  in  time, 
 $$
\lann f \rann = \lann f(t) \rann= {\mathbb E}  \, \frac1{\sigma} \int_T^{T+\sigma} f(s) \,ds, 
$$ 
 where  $ T\ge1$ and $\sigma\ge \sigma_*$ are parameters of the averaging. 
   In this notation the just proved result reeds 
 $
\lann \| u^\nu\|_1^2\rann \ge \nu^{-1} \,\tfrac12 B_0.
$
But by  Theorem  \ref{main_est}   $\lann \| u^\nu\|_1^2\rann \le \nu^{-1} C$.  So  
$$
\lann \| u^\nu \|_1^2\rann \sim \nu^{-1}, 
$$
where $\sim$ means that the ratio of the two quantities is bounded from below and from above, uniformly in $\nu$ and in the parameters 
$T\ge1$ and $\sigma\ge \sigma_*$,
entering the definition of the brackets $\lann \cdot  \rann $.

Now the 
 Gagliardo-Nirenberg   inequality  jointly with \eqref{OK}  imply:
 $$
\lann | u_x^\nu|_{L_2}^2\rann
\le C'_m \lann \| u^\nu\|_m^2\rann^{\frac{1}{2m-1}}  \lann | u^\nu_x|_{L_1}^2\rann^{\frac{2m-2}{2m-1}} \le
 C_m \lann \| u^\nu\|_m^2\rann^{\frac{1}{2m-1}} , \quad m\in \N. 
$$
Using the already obtained  lower bound for the  first Sobolev norm
 we get from here  lower bounds for the second moments of  all norms  $ \| u^\nu\|_m$:
 $$
\lann \| u^\nu\|_m^2\rann \ge C''_m \nu^{-(2m-1)} \qquad \forall m\in \N. 
$$
Combining this with the upper bound in Theorem \ref{main_est} we get:

\bt \label{t_Sob_norm}
 For any $u_0\in H^1$, any $0<\nu\le1$   and every $m\in\N$ ,
     \begin{equation}  \label{up_down}
      \lann \| u^\nu (t;u_0) \|_m^2\rann  \sim   \nu^{-(2m-1)} .
  \ee
  \et

This theorem and the Oleinik estimate  turn out to be a powerful and 
 efficient tool to study  turbulence in the 1d Burgers equation \eqref{B} (the burgulence).  In particular, they imply that 
 $$
 \llan  \|u^\nu(t;u_0)\|_m^2 \rran \sim 1 \qquad \forall u_0 \in H^1, \;\forall\, m\le0.
 $$
 Here the upper bound $ \llan  \|u^\nu(t;u_0)\|_m^2 \rran  \le C^{-1}$ for $m\le0$  immediately follows from \eqref{OK}, 
 while derivation of the lower estimate  $ \llan  \|u^\nu(t;u_0)\|_m^2 \rran  \ge C_m^{-1}$  for $m\le0$ 
 requires some efforts.

We stress that we do not know if  $  \lann \| u^\nu\|_m^2\rann$ admits an asymptotic expansion as $\nu\to0$, i.e. if it is true that 
 $$
   \lann \| u^\nu\|_m^2\rann = C_m  \nu^{-(2m-1)}  +o( \nu^{-(2m-1)} ), \qquad m\in \N, 
 $$
 for a suitable constants $C_m$.

\section{Burgulence and K41
}\label{s_3}

\subsection{Dissipation scale }

  By a direct analogy with K41, the     basic quantity, characterizing a solution $u^\nu(t,x)$  of  \eqref{B} 
     as a 1d turbulent flow is its {\it dissipation scale}  $l_d$, a.k.a. 
      {\it Kolmogorov's inner scale}. To define the {\it mathematical dissipation scale} $l_d(u)$  of any random field $u^\nu(t,x)$ which 
      depends on a parameter $\nu\in(0,1]$ and   defines a 
      random process $u^\nu(t,\cdot) \in L_2$,  we write $u^\nu$  as Fourier series 
      $
        u^\nu(t,x) =\sum \hat u_k^\nu(t) e^{2\pi ikx}.
      $
      Then we 
     set $l_d$ to be the smallest number 
      of the form $l_d= \nu^{-c_d}$, $c_d>0$, such   that  for $|s| \gg l_d$ the averaged 
    squared norm of the  $s$-sth Fourier coefficient $\hat u^\nu_s(t)$ decays with $s$     very fast. Namely, $c_d$ is the smallest positive  number 
    with the property that for each $\gamma>0$ 
   $$
   \lann | \hat u^\nu_s(t)|^2\rann \le C_{N,\gamma}  |s|^{-N} \quad \text{if} \;\; \;
    |s| \ge  \nu^{-c_d-\gamma}.
   $$
    If such a $c_d>0$   does not exist, then the inner scale  $l_d(u)$ is not defined.  If
   $u$ does not depend on $t$ or is stationary in $t$, 
   then, naturally, in the relation above the the averaging $\lann \cdot \rann$ may be replaced by $\EE$.

 Theorem  \ref{t_Sob_norm}  and estimates  \eqref{conseq} with $p=2$    imply:
 
 \bt\label{t_diss_scale}
   The  mathematical  dissipation  space-scale $l_d$  of any solution 
  $u$ of eq. \eqref{B}   with $u_0\in H^1$  equals $\nu^{-1}$.
  \et

  In physics, the  dissipative scale  $l_d$ is defined modulo a constant factor, so for the Burgers equation the physical dissipative scale is 
$l_d=$Const$\,\nu^{-1}$.  It was  Burgers himself who first   predicted the  correct   value of $l_d$.

Now let us consider the set of integers $[C_1, \infty)$, regarded as the set of indices  $s$ of Fourier coefficients $\hat u_s$, 
and the closed interval $[0, c_1]$,  $c_1\le1/2$, regarded as the set of increments of $x$. 
 Using the physical dissipative scale  $l_d$  we divide both of them to two sets, called the {\it 
dissipation  and  inertial ranges}:\footnote{ In this paper we do not deal with the energy range, so we do not define it.} 
 
 \noindent 
 -- in Fourier presentation the {\it dissipation range} is $I_{diss} = (l_d, \infty) = (C\nu^{-1}, \infty)$, and the 
  {\it inertial  range} is $I_{inert} = [\,$const$,  l_d] = [ C_1, C\nu^{-1}] $. 
  
  \noindent 
  -- in the $x$-presentation the  { dissipation range} is  $I^x_{diss} = [0, c \nu)\subset [0, 1/2]$, and  the 
  { inertial  range} is $I^x_{inert} = [c \nu,  c_1]\subset [0, 1/2]$. 
  
   The constants $C, C_1$ and $c,c_1$  do not depend on $\nu$ and may change from one group of results to another.

\medskip
\noindent {\it Dissipation scale in K41.}
  In K41 the hydrodynamical dissipation scale is predicted to be  $l_d^K =$Const$\,\nu^{-3/4}$. Accordingly, in the Fourier presentation the inertial 
  range of the K41 theory    is $ I^K_{inert}=
  [C_1, C\nu^{-3/4}]$, while in the $x$-presentation it is $ I^{xK}_{inert}=
  [c \nu^{3/4}, c_1 ]$.

 \subsection{Moments of  small-scale increments}
 
 For a random field $u=u(t,x)$, $t\ge0,\,  x\in S^1$, such that $u(t,\cdot)$ is a random process in $L_p$ for every $p< \infty$, 
  we consider the moments of its space-increments, average them in $(x, t)$ and organise the result in the { \it
 structure function of $u$}:
 \be\label{Sp}
 S_{p,l}(u) = \lann | u(\cdot+l) -u(\cdot ) |^p_{L_p} \rann, \quad p>0, \; \; | l| \le 1/2.
 \ee

 If $u= u^\nu(t,x;u_0)$  with some $u_0\in H^1$, then a.s.  for $t>0$   $u$ is a smooth function of $x$, so for very small $l$ the 
 function   $S(u)$  behaves  as $| l|^p$. It turns out that for $l$ not that small it behaves differently:

 \bt\label{t_small_scale}
 For $u=u^\nu$ as above  and for    $|l|$ in the inertial range   $ [c \nu, c_1 ]$  we have  
\be\label{inertial_scale}
S_{p,l}(u^\nu) \sim |l|^{ \min(p,1)} \quad \forall\, p>0. %\\&S_{p,l}(u^\nu) \sim | l | \quad \text{if}\quad  p\ge1.
\ee
While for $|l|$ in  the dissipation range  $ [0, c \nu)$, 
 \be\label{diss_scale}
 S_{p,l}(u^\nu) \sim|l|^p \nu^{1-\min(p,1)} \quad   \forall\, p>0. 
 \ee
 The constants $c$ and $ c_1$  depend only on the force \eqref{force}. 
 \et

 In \cite{Fr1} U.~Frisch with collaborators obtained  the  assertion \eqref{inertial_scale}  by a convincing heuristic argument. 
We rigorously derive  \eqref{inertial_scale}  and  \eqref{diss_scale}  from Theorems~\ref{OK} and 
\ref{t_Sob_norm},  using some ideas from \cite{Fr1}.

\medskip
\noindent {\it Moments of  small-scale increments in K41.}
 For  water turbulence  the structure function is defined as above with the difference that  there 
 the increment of the velocity field $u(x+r) - u(x)$ (usually) is
 replaced by its projection on the direction of the vector $r$. Since the K41 theory deals with stationary and homogeneous vector fields, then there
 the structure function of a velocity field $u(t,x)$, $x\in \T^3$,   is defined as 
 \be\label{K_str}
 S^{\Vert}_{p,r} (u) = \EE\Big |\big (u(t,x+r) - u(t,x) \big) \cdot \frac{r}{|r]} \Big|^p
 \ee
 (the r.h.s. does not depend on $t$ and   $x$, and we recall that in K41  \    $\EE u(t,x) \equiv0$).  
 The Kolmogorov theory predicts that if the viscosity of the fluid is    $\nu\ll1$ (so  the Reynolds number  is large), then  
\begin{equation} \label{p2}
S^{\Vert}_{2,r}(u)
{\sim} |r|^{2/3} \quad\text{  for $|r|  \in I_{inert}^{xK}$
}.
\end{equation}
This is the celebrated   {\it 2/3 law of the  K41 theory}. 
The theory  states that the third moment of the speed's  increments without the modulus sign behaves similarly:\footnote{ This
 relation implies that the random field $u$ is not Gaussian since for Gaussian fields the l.h.s. of \eqref{p3} vanishes.}
\begin{equation} \label{p3}
\left\langle\big( {(u(x+r)-u(x)) \cdot r/|r|}\big)^3\right\rangle  \sim -r \quad\text{  for $|r|  \in I_{inert}^{xK}$
}.
\end{equation}
The dimension argument, used by Kolmogorov to derive \eqref{p2},  also  implies that 
\begin{equation} \label{p>3}
S^{\Vert}_{p,r}(u)
{\sim} |r|^{p/3}\quad \;\; \text{for \; $ p > 0$  if\; \;$|r|  \in I_{inert}^{xK}$}.
\end{equation}
This relation, although not claimed in the K41 papers, was frequently suggested  in later works, related to K41. 
 \medskip

\medskip
\noindent {\it Burgulence compare to   K41.} In \eqref{p2},  \eqref{p>3}  the structure function  behaves as $| r |$, raised to a 
degree, proportional to $p$, while in \eqref{inertial_scale} the degree is a nonlinear function of $p$. Based on that  the relation in
\eqref{inertial_scale} sometime is called the abnormal scaling. The linear in $p$ behaviour of the exponent in \eqref{p>3} now 
is frequently put to doubt. Indeed, it implies that for any $p,q>0$  the ratio 
      \be\label{g_like}
      {(S^{\Vert}_{p,r})^{1/p} }\big/ {(S^{\Vert}_{q,r})^{1/q}}  \sim C_{p,q} \quad\text{  for $|r|  \in I_{inert}^{xK}$\,,}
       \ee
       where $C_{p,q}$ 
       is an $r$-independent quantity. If $u(x+r) -u(x) =:\zeta$ was a Gaussian r.v., then the 
       relations \eqref{g_like} would hold as equalities    with absolute constants  $C_{p,q}$,
       independent from $\zeta$.  But it is well known from experiments (and follows from \eqref{p3})      that  increments 
          of  the velocity field $u$  of a fluid  with small viscosity 
           are not Gaussian, so the Gaussian-like behaviour, manifested by \eqref{g_like}, looks suspicious. On the 
          contrary, if $u=u^\nu(t;u_0)$ is a solution of \eqref{B}, then  in view of \eqref{inertial_scale}, for $p,q\ge1$ we have
      $$
      {S_{p,r}^{1/{p}} } \big/{S_{q,r}^{1/{q}} }  \sim C_{{p},{q}} \,|r|^{1/{p} -1/q} \quad\text{  for $|r|  \in I_{inert}^{x}=[c\nu, c_1]$,}
       $$
       which is big if $p>q$ and $|r| \in I^x_{inert}$ is small (the latter  may be achieved if $\nu\ll1$). 
        This  very non-Gaussian behaviour\,\footnote{The function above with $p=4,\, q=2$          is called the 
         {\it flatness } of the random variable $u(x+l) -u(x)$. It equals  three  for any  Gaussian r.v.} 
         of the increments of $u$  shows that solutions of 
         \eqref{B} with small $\nu$ are random fields, far from Gaussian.

   \subsection{Distribution of energy along the spectrum}\label{s_3.3} 
  The second celebrated law of the Kolmogorov theory deals with the distribution of  fluid's  energy
   along the spectrum.  
   
     For a random field   $u(t,x)$ which defines a random process $u(t, \cdot) \in L_2$ we define its energy as $\tfrac12 \lann |u|_{L_2}^2\rann$ 
     (this is a  common convention). By Parseval's identity, 
      $$
      \Big \langle\!  \Big\langle  \tfrac12  \int |u|^2dx \Big \rangle\!  \Big\rangle
       = \sum_s \tfrac12 \lann |\hat u_s|^2\rann, 
       $$
      so the quantities $\tfrac12 \lann |\hat u_s|^2\rann$ characterize  distribution of  energy of the field $u$      along the spectrum.
  Next, for any $k\in\N$ we define $E_k(u)$   as the averaging of  $\tfrac12 \lann |\hat u_s|^2\rann$  along the layer $J_k$ 
  around $k$, 
  $$
  J_k =\{ n\in \Z^* : M^{-1} k \le |n| \le M k\}, \quad M>1.
  $$
  I.e., 
  \be\label{Ek}
  E_k(u)= \lann e_k(u)\rann, \qquad e_k(u) = \frac1{| J_k|} \sum_{n\in J_k} \tfrac12 |\hat u_n|^2.
  \ee
  %where   
  %$$ e_k(u)  = \frac1{| J_k|} \sum_{n\in J_k} \tfrac12 |\hat u_n|^2$$
  % is the averaged energy of the $k$-th mode of $u$.  
  The function $k\mapsto E_k(u)$ is called the {\it energy spectrum} of the random field 
  $u$.  
  
  If $u=u^\nu(t;u_0)$ is a solution of \eqref{B}, then it follows immediately from the definition of $l_d(u)$  that      for $k\gg l_d$ \ $E_k(u)$ decays 
  faster than any negative degree of $k$, uniformly in $\nu$. But for $k \le l_d$ the behaviour of $E_k$ is quite different. Namely, 
 Theorem~\ref{t_small_scale} and relations 
  \eqref{conseq}  imply the  following spectral power law for  "1d Burgers fluid":
 
 \bt\label{t_power_law}
 Let $u$ be a solution of eq.~\eqref{B} with  any $u_0\in H^1$. Then for  $k$ in the inertial range, 
  $1\le k\le C \nu^{-1}$, we have:
  \begin{equation}\label{power} 
  E_k (u^\nu) \sim  k^{-2}, %\quad \forall\, \nu>0, \quad \qquad\quad \qquad \qquad \qquad   (*)
\ee
with suitable $C>0$ and $M >1$, depending only on the random force. 
\et

 For solutions of \eqref{B}, Jan  Burgers already in 1940  predicted that  $E_k \sim | k|^{-2}$ for $|k| \le\, $Const$\, \nu^{-1}$, 
 i.e. exactly the  spectral power law above. 
 
 We do not know if the theorem's assertion remains true for any $M>1$ (with a suitable $C(M)$). 
 
 Let us briefly explain how  \eqref{power}  follows from Theorem~\ref{t_small_scale}. For a solution 
 $u=u^\nu(t,x; u_0)$ relation \eqref{conseq} implies the upper bound for  energy spectrum, 
 $
 E_k (u^\nu) \le C  k^{-2}
 $
 for each $k$,  as well as that 
 
\begin{equation} \label{as_well}
\sum_{|n|\leq M^{-1}k}|n|^{2}\langle\langle|
\hat u_{n}|^{2}\rangle\rangle\leq CM^{-1}k, \qquad 
\sum_{|n|\geq Mk}\langle\langle|
\hat u_{n}   |^{2}\rangle\rangle\leq C^{\prime}M^{-1}k^{-1}.  
\end{equation}
Now consider the sum 
$\Sigma_k=\underset{|n|\leq Mk}{\sum}|n|^{2}\langle\langle|
\hat u_{n} |^{2}\rangle\rangle.$ Since $|\alpha| \ge |\sin \alpha | $, then
$$
 \Sigma_k\geq %\frac{k^{2}}{\pi^{2}}\sum_{|n|\leq Mk}\sin^{2}(n\pi k^{-1} )\langle\langle|
%\hat u_{n} |^{2}\rangle\rangle \\ \nonumber
\frac{k^{2}}{\pi^{2}}\Big(  \sum_{n=-\infty}^\infty
%\mathbb{Z}^{\ast}}
\sin^{2}(n\pi k^{-1})\langle\langle|
\hat u_{n} |^{2}\rangle\rangle-\sum_{|n|>Mk}\sin^{2}(n\pi k^{-1})\langle\langle|
\hat u_{n} |^{2}\rangle\rangle\Big)  .
$$
By Parseval's identity, 
$
|
u(t,\cdot+y)-
u(t,\cdot)|^{2}_{L_2}=4\sum_{n\in
\mathbb{Z}
^{\ast}}\sin^{2}(n\pi y)|
\hat u_n(t)|^{2}.
$
Applying the averaging $\llan\cdot\rran$ to this equality we get that the structure function 
$
S_{2, 1/k} (u)$  equals  to 
 $  4\sum_{n} \sin^{2}(n\pi k^{-1} )\llan\hat u_n(t)\rran^{2}.
$
So 
\begin{equation*}
\Sigma_k\geq\frac{k^{2}}{\pi^{2}}\Big( \frac14 S_{2, 1/k} (u) 
%\frac{1}{4}\langle\langle\| u(\cdot+k^{-1})-
%u(\cdot)\|^{2}\rangle\rangle
-\sum_{|n|>Mk}\langle\langle|
\hat u_{n}|^{2}\rangle\rangle\Big). % \geq k^{2}C_{1}S_{2,k^{-1}}(u) -C'M^{-1}k. 
\end{equation*}
Using the second inequality in \eqref{as_well} and Theorem \ref{t_small_scale} we find that 
$
\Sigma_k \ge k^2 C_1 k^{-1} - C_2 M^{-1} k. 
$
Since 
\[
E_{k}\geq\frac{1}{2k^{3}M^{3}}\Big(  \Sigma_k -\sum_{|n|\leq M^{-1}k}|n|^{2}
\langle\langle|
\hat u_{n}|^{2}\rangle\rangle\Big),
\]
then using the just obtained lower bound for $\Sigma_k$ and the first inequality in  \eqref{as_well} 
we get that $E_k \ge C^{-1} k^{-2}$, if $M$ is large enough. 
 
\medskip
\noindent {\it  Distribution of  energy along the spectrum in K41}.
  For the water turbulence the    K41 theory  predicts that $E_k$ obeys the celebrated  Kolmogorov--Obukhov law: 
  \begin{equation}\label{KOb}
 E_k \sim | k|^{-5/3} \quad \text{ for \  $k$ \  in the inertial range. }
 \ee
 Experiments and numerical study of the corresponding equations convincingly  show that this law is close to reality, see \cite[Section~5.1]{F}.

  \subsection{Relation between the two laws of turbulence.}\label{s_3.4} 
  Let us first note that the definitions of the structure function $S$ and    the energy spectrum $E_k$ of a random field $u$ 
   in Section~\ref{s_3}  apply 
  in the case when $u=u(x)$ does not depend on $t$ and $\om$. 
  Then the averaging $\llan\cdot\rran$ may be dropped in the definitions of the objects.    In this case
  the proof of Theorem~\ref{t_power_law}, sketched in  Section~\ref{s_3.3}, shows that if  a function 
  $u^\nu(x) \in H^1$, depending  on  $\nu\in(0,1]$,
  is normalised by the relation 
  $
  |u^\nu|_{L_2} \equiv 1
  $
 and for all $\nu$  satisfies
  
  1) relation \eqref{inertial_scale} with $p=2$ for $|l| \in [ c \nu, c_1]$, 
  
  2) relation \eqref{conseq}, which for $u=u(x)$  reeds 
  $
  | \hat u^\nu_k| \le C |k|^{-1}
  $
  for all $k$,
  \smallskip
  
  \noindent
  then the assertion of Theorem~\ref{t_power_law} holds with a suitable $C$ and a sufficiently big 
   $M$ (certainly same is true if $u^\nu$ is a random field). 
  
   It is very likely (but we have not checked 
  this) that, on the contrary, the assertion of Theorem~\ref{t_power_law} jointly with relation  \eqref{conseq} (or \eqref{OK}), which should be understood 
  as in 2) above, imply the validity of \eqref{inertial_scale} for  $|l| \in [ c \nu,  c_1]$ with suitable $c, c_1>0$, 
  and for  $p=2$  (probably also for all $p>0$). 
  \medskip
  
  Much more interesting and more involved is  the relation between the 2/3-law \eqref{p2} with 
  $|r|  \in I_{inert}^{xK}$ and Kolmogorov--Obukhov law
  \eqref{KOb} with $ |k| \in I_{inert}^K$, for any 3d random  field $ u(x)$, depending on a parameter $\nu$. 
  On a physical level of rigour it is explained on pp.~134-135 of 
  \cite{LL} that the two laws are equivalent for sufficiently general fields $u$, but for a mathematical reader this explanation 
  seems rather insufficient.\,\footnote{The corresponding argument was added by E.~Lifschitz to
  the third Russian edition of the book, after L.~Landau passed away. In that version of the book (which corresponds  to the second English edition 
  \cite{LL}) the  part, dedicated to the theory of turbulence, was significantly edited and enlarged.  }
  In \cite[$\S 21.4$]{MY} (also see \cite[Section~4.5]{F}), assuming that $u(x)$ is an homogeneous and
   isotropic random field on $\R^3$, the    equivalence of the two laws is established by a formal calculation, 
   based on  the spectral representation for $u(x)$ (see \cite[\S11.2]{MY}).  By analogy with what was said above concerning 
    the 
   two laws of burgulence, it seems that this calculation cannot be rigorously justified without imposing  additional 
   restrictions on $u(x)$ (and/or   its  Fourier transform),  cf. above assumption 1).  So we think that   without referring 
   to some additional   properties of fluid's  flow with large Reynolds number (e.g. without evoking a 
     new estimate for solutions of the 3d Navier--Stokes system),  
     the two laws of turbulence should  be regarded not as the same assertion, written in the 
   $x$\,- and in Fourier presentations, but rather as two different (although related) statements. 
  
  To find a natural sufficient condition which would guarantee for a vector field $u^\nu(x)$ 
  on $\T^3$ (or for a stationary field on $\T^3$, or on 
  $\R^3$) equivalence of the two laws of the K41 theory, or at least that one of them implies another, is an interesting 
  open question. The field $u^\nu$ should be normalise by the relation
   $
  |u^\nu|_{L_2} \equiv 1,
  $
  or 
   $
 \EE  |u^\nu(x)|^2 \equiv 1
  $
  if it is a  stationary field on $\T^3$. If $u^\nu$ is a stationary field on $\R^3$, it should be assumed that its correlation uniformly in $\nu$ is
  a tensor of order one, fast decaying at infinity. 

   The technique, developed to prove the equivalence of the two laws under a hidden additional   condition, may allow to 
  calculate the asymptotic  of    $S^{\Vert}_{p,r}(u)$ for $p>0$ and 
   $|r|$ in the inertial range, and thus to correct relation \eqref{p>3}, which most likely 
  is wrong for large $p$.

  \section{The mixing} The mixing  in eq. \eqref{B} means    that  in a function space $H^m, \, m\ge1,$ 
  where we study the  equation,    there exists a unique Borel  measure $\mu_\nu$, such that for 
  any ``reasonable" functional $f$ on $H^m$  and for any solution $u(t,x;u_0)$, $u_0\in H^1$, 
   we have 
\be\label{mix}
{\mathbb E} f(u(t;u_0)) \to \int_{H^m} f(u)\, \mu_\nu(du) \quad \text{as} \quad t\to\infty. 
\ee
The measure $\mu_\nu$ is called 
the stationary measure for eq. \eqref{B}. If $u_0$ is a r.v., distributed as $\mu_\nu$, 
then $u(t;u_0) =: u^{st}(t)$ is a stationary solution: $\cD ( u^{st}(t)) \equiv \mu_\nu$. 

It  may be derived from a general theory that the mixing 
 holds for eq.~\eqref{B}, but  then the rate of  convergence in \eqref{mix}  would 
depend on $\nu$. In the same time,  in the theory of turbulence the rate of convergence to a statistical equilibrium 
 should not depend on the viscosity  (see in \cite{Bat}, e.g. pages 6-7 and 109), 
 and  for solutions of  \eqref{B} it does not:

  \bt\label{t_mixing}
 If the functional $f(u)$ is continuous in  some  $L_p$--norm, $p<\infty$,  and 
 $
 |f(u)| \le C |u|^N_{L_p}$ for suitable $C,N>0$, 
 %and $N\in \N$, 
  then \eqref{mix} holds. The   rate of convergence is  at least  $(\ln t)^{-\kappa_p}$, for some $\kappa_p>0$. 
\et

The proof follows from the results in Section \ref{s_2}, basic methods to prove the mixing in stochastic PDEs,  and from 
another remarkable feature of the Burgers equation: %if $\om\in\Omega$ belongs to the full-measure set  for which the 
%force $\xi^\om(t,x)$ is well defined, then for any $u_0$ and $ u_1$, 
$$
\big| u^\om(t;u_0) - u^\om(t;u_1) |_{L_1} \le |u_0-u_1|_{L_1} \quad \text{for  every}\;\; t\ge0,
$$
for a.a. $\omega$ and all $u_0, u_1 \in H^1$.

 If in \eqref{force}  $b_s\equiv b_{-s}$, then the random field $\xi(t,x)$ is 
homogeneous in $x$. In this case the measure $\mu_\nu$ also is homogeneous, as well as 
 the stationary solution $u^{st}(t,x)$. All results in Section~\ref{s_3} remain true for $u^{st}$,  which describes
   the stationary and space-homogeneous burgulence.

Energy spectrum of the stationary measure $ \mu_\nu $  is
$
E_k(\mu_\nu) = \int e_k(u) \mu_\mu(du),  %\qquad e_k(u ) = \frac1{| J_k|} \sum_{n\in J_k} \tfrac12 |\hat u_n |^2.
%{\mathbb E} f(u(t, \cdot)) \to \int_H f(u)\, \mu_\nu(du) \quad \text{as} \quad t\to\infty. 
$
where $e_k$ is as in \eqref{Ek}. 
Obviously, 
  $$
  E_k(\mu_\nu)= \lann e_k(u^{st} (t))\rann  =  \EE  e_k(u^{st}(t)).
   $$
   Since $\lann e_k(u^{st} (t))\rann$ satisfies the spectral power law, then $E_k(\mu_\nu)$ also does:
  $$
  E_k(\mu_\nu) \sim k^{-2}\quad\text{ for $1\le k\le C \nu^{-1}$.
  }
   $$
   Due to Theorem \ref{t_mixing} the instant energy spectrum of every solution converges to that of $\mu_\nu$: 
   $$
   \EE e_k(u(t;u_0)) \to E_k(\mu) \qquad \forall\, u_0 \in H^1,\; \forall\, k\in\N, 
   $$
   uniformly in $\nu$.

    Similarly the structure function of $\mu_\nu$,  defined as 
    $
    S_{p,l}(\mu_\nu) =  \int_{H^m} | u(\cdot +l)- u(\cdot) |^p_{L_p} \mu_\nu(du),
    $
    satisfies \eqref{inertial_scale} and \eqref{diss_scale} for $l$ in the inertial and dissipation ranges, correspondingly.  As above, the instant structure function 
    of every solution converges, as time grows,      to $S_{p,l}(\mu_\nu)$ for  all $p$ and $l$, uniformly in $\nu$ (and in $l$).       If
    $b_s\equiv b_{-s}$, then the measure $\mu_\nu$ is homogeneous and then 
    $$
    S_{p,l}(\mu_\nu) = \EE | u^{st}(t,x+l) - u^{st}(t,x)|^p  =\!  \int_{H^m} | u(x +l)- u(x) |^p \mu_\nu(du)
    \quad \text{for any} \; \; t,x. 
    $$
   
\smallskip
\noindent {\it  Mixing and the theory of turbulence}.
      The results in this section are in line with the general theory of turbulence which postulates that statistical characteristics of turbulent flows converge, as time grows, 
    to a universal statistical equilibrium. They also are in the spirit of K41, where the velocity field of a fluid     is assumed to be   stationary in $t$ and homogeneous in $x$.

\section{Inviscid limit}
 Another remarkable feature of the Burgers equation \eqref{B} is that, as $\nu\to0$ (so the Reynolds number of the corresponding ``1d fluid" grows to infinity),
  the  solutions of the equation    converge to   inviscid  limits:  
$$
u^\nu(t, \cdot;u_0) \rightarrow u^0(t, \cdot;u_0) \quad \text{ in \ $L_p(S^1)\; \;\forall\, t\ge0$, \ a.s.}, 
$$
for every  $p<\infty$ and every $u_0$. 
This result 
is due to Lax--Oleinik (1957). The limit $u^0(t,x;u_0) $ is called an ``inviscid solution", or an ``entropy solution" of equation 
 \eqref{B} with $\nu=0$.  The limiting  function $u^0(t,x;u_0)$ is bounded in $x$ for every $t$, but in general is  not continuous.  Still its 
  structure function and spectral energy are well defined  and   inherit the laws, proved  for $u^\nu$ with $\nu>0$. For $u^0$ 
  the laws are valid 
  with $\nu=0$ in their statements: 
  
  \bt\label{t_inviscid}
   For each entropy solution $u^0$, 
   
 1) $ E_k(u^0) \sim k^{-2} $ for all $k$;
 
 2) $S_{p,l}( u^0) \sim|l|^{\min(p,1)} \;
 \;$ if $p>0$ and %  $|l| \le {C_1}$, 
% and 
 % ${}\; S_p(l, u^0) \sim |l| \;$ if $\;p\ge1$
  $|l| \le {c_1}$.
\et

Since  the  spectral power law  for $E_k(u^0)$ holds for all $\ k\ge1$, then the dissipation  range of $u^0$  is empty.
Its  inertial range  in the Fourier presentation is the interval $[1,\infty)$,  and in $x$ --  the  interval
$[0, c_1]$. The inviscid solutions define in the space
$
L_1=\{ u\in L_1(S^1):\int u\,dx=0\}
$
a mixing Markov process, whose  stationary measure is supported by the space 
$
L_1\cap\big( \cap_{p<\infty} L_p(S^1)\big). 
$
\medskip

These results   describe  the {\it inviscid burgulence}. They have  no analogy in  the K41 theory since there the Reynolds number $Rey$ of  fluid's flow 
is a fixed finite quantity, and since on the mathematical side of the question, behaviour of solutions of the 3d hydrodynamical equations on time-intervals 
of order $\gsim1$ when $Rey \to \infty$ is a completely open problem.

\section{Conclusions} 
 The stochastic Burgers equation \eqref{B}   with small viscosity  makes a consistent model of 1d turbulence. Its
 rigorously proved statistical properties make natural and close analogies for the main laws of the K41 theory of turbulence. This, once again, 
 supports the belief that the K41 theory is ``close to the truth".

\bibliographystyle{amsplain}
\bibliography{Biblio}

\end{document}